\documentclass[longauth,traditabstract]{aa}  
\usepackage{graphicx}
\usepackage{txfonts}
\usepackage{subfigure}
\usepackage{amssymb}
\usepackage{url}

\def\kms{\rm {km\,s$^{-1}$}}

\def\etal{et~al.}
\def\cmsq{\rm {cm$^{-2}$}}
\def\cmcub{\rm {cm$^{-3}$}}

\def\h2o{H$_2$O}
\def\h2o18{H$_2^{18}$O}
\def\so2{SO$_2$}
\def\13co{$^{13}$CO}
\def\NH3{NH$_{3}$}
\def\nh2{NH$_{2}$}
\def\hco+{HCO$^{+}$}
\def\water18{H$_2^{18}$O}

\def\g10{G10.6$-$0.4}

\begin{document}
%
\title{Nitrogen hydrides in interstellar gas}
  \subtitle{{\emph{Herschel}}\thanks{\emph{Herschel} is an ESA space observatory with science instruments provided by European-led  Principal Investigator consortia and with important participation from NASA.}/HIFI  observations  towards G10.6-0.4 (W31C)} 
\authorrunning{C.M.~Persson \etal}   
   \author{C.M.~Persson  
          \inst{1},
          J.H.~Black\inst{1},
	J.~Cernicharo\inst{2}, J.R.~Goicoechea\inst{2},
	 G.E.~Hassel\inst{3}, 
E.~Herbst\inst{4},    
	M.~Gerin\inst{5},   M.~De~Luca\inst{5},    
	T.A.~Bell\inst{6},  
	A.~Coutens\inst{7,8},
        E.~Falgarone\inst{5}, 	
	 P. F.~Goldsmith\inst{9}, 
	  H.~Gupta\inst{9},
	  M.~Ka{\'z}mierczak\inst{10}, 
	D.C.~Lis\inst{6},
	B.~Mookerjea\inst{11},
	D.A.~Neufeld\inst{12},
	J.~Pearson\inst{9}, 
	T.G.~Phillips\inst{6}, 
	P.~Sonnentrucker\inst{12},
	J.~Stutzki\inst{13},  
	C.~Vastel\inst{7,8}, 
	S.~Yu\inst{9}, 
F.~Boulanger\inst{14},  	  
  E.~Dartois\inst{14},
 P.~Encrenaz\inst{5}, 
 T.R.~Geballe\inst{15}, 
 T.~Giesen\inst{13},
B.~Godard\inst{5},
C.~Gry\inst{16}, 
 P.~Hennebelle\inst{5}, 
P.~Hily-Blant\inst{17},
C.~Joblin\inst{7}, 
R.~Ko{\l}os\inst{18}, 
J.~Kre{\l}owski\inst{10}, 
J.~Mart{\'\i}n-Pintado\inst{2}, 
K.~Menten\inst{20}, 
R.~Monje\inst{6}, 
M.~Perault\inst{5},   
R.~Plume\inst{21}, 
M.~Salez\inst{5},   
S.~Schlemmer\inst{13}, 
M.~Schmidt\inst{19}, 
D.~Teyssier\inst{22}, 
I.~P\'{e}ron\inst{5, 23}, 
P.~Cais\inst{24}, 
P.~Gaufre\inst{24},  
A.~Cros\inst{7,8},     
L.~Ravera\inst{7,8},     
P.~Morris\inst{26},         
S.~Lord\inst{26}, 
P.~Planesas\inst{27}
          }   
   \offprints{carina.persson@chalmers.se}
  \institute{Onsala Space Observatory, Chalmers University of Technology, SE-439 92 Onsala, Sweden\\  
             \email{\url{carina.persson@chalmers.se}}
\and Centro de Astrobiolog\`{i}a, CSIC-INTA, 28850, Madrid, Spain 
\and   Depts. of Physics,  Ohio State Univ. USA
\and   Depts. of Physics, Astronomy \& Chemistry, Ohio State Univ. USA
\and LERMA, CNRS, Observatoire de Paris and ENS, France 
\and California Institute of Technology, Cahill Center for Astronomy and Astrophysics 301-17, Pasadena, CA 91125, USA  
\and  Centre d'etude Spatiale des Rayonnements (CESR), Universite de Toulouse [UPS], 31062 Toulouse Cedex 9, France and  
\and  CNRS/INSU, UMR 5187, 9 avenue du Colonel Roche, 31028 Toulouse Cedex 4, France   
\and  JPL, California Institute of Technology, Pasadena, USA
\and   Nicolaus Copernicus University, Toru{\'n}, Poland 
\and   Tata Institute of Fundamental Research, Homi Bhabha Road, Mumbai 400005, India
\and  The Johns Hopkins University, Baltimore, MD 21218, USA 
\and  I. Physikalisches Institut, University of Cologne, Germany 
\and   Institut d'Astrophysique Spatiale (IAS), Orsay, France 
\and  Gemini telescope, Hilo, Hawaii, USA 
\and  LAM, OAMP, Universit\'{e} Aix-Marseille \& CNRS, Marseille, France
\and   Laboratoire d'Astrophysique de Grenoble, France
\and   Institute of Physical Chemistry, PAS, Warsaw, Poland 
\and Nicolaus Copernicus Astronomical Center, Toru{\'n}, Poland 
\and  MPI f\"ur Radioastronomie, Bonn, Germany
\and   Dept. of Physics \& Astronomy, University of Calgary, Canada
 \and  European Space Astronomy Centre, ESA, Madrid, Spain 
\and   Institut de Radioastronomie Millimetrique, IRAM, 300 rue de la Piscine, F-38406 St Martin d'Heres 
\and  Institute Universite de Bordeaux, Laboratoire d'Astrophysique de Bordeaux, 33000 Bordeaux, France and  CNRS/INSU, UMR  5804, B.P. 89, 33271 Floirac cedex, France 
\and    Centre d'etude Spatiale des Rayonnements (CESR), Universite de Toulouse [UPS], 31062 Toulouse Cedex 9, France   
\and   Infrared Processing and Analysis Center, California Institute of Technology, MS 100-22, Pasadena, CA 91125  
\and    Observatorio Astron\' Nacional (IGN) and  Atacama Large Millimeter/Submillimeter Array, Joint ALMA Office, Santiago, Chile  
}

   \date{Received May 31, 2010 /  Accepted July 6, 2010}

 
  \abstract
{  
The HIFI instrument on board the \emph{Herschel} Space Observatory has been used to observe interstellar 
nitrogen hydrides   along the sight-line towards  \mbox{\g10}~in order to improve our understanding
of the interstellar chemistry of nitrogen.
We report observations of absorption in NH 
\mbox{$N\!=\!1\leftarrow0$},  \mbox{$J\!=\!2\leftarrow1$} and  $ortho$-\nh2~\mbox{$1_{1,1} \gets 0_{0,0}$}. 
We also observed   $ortho$-\NH3 \mbox{$1_0 \gets 0_0$}, and  \mbox{$2_0 \gets 1_0$},   $para$-\NH3~
\mbox{$2_1 \gets 1_1$}, and searched unsuccessfully for NH$^+$. 
All detections show emission and absorption associated directly with
the hot-core source itself as well as absorption by foreground material over a wide range of velocities.
All spectra show similar, non-saturated, absorption features, which we attribute to diffuse molecular
gas. 
Total column densities over the velocity range 11\,--\,54\,\kms~are estimated.
The similar profiles suggest fairly uniform abundances relative to hydrogen, approximately  
$6\times 10^{-9}$,  $3\times 10^{-9}$, and $3\times 10^{-9}$ for NH, \nh2, and \NH3, respectively.
These abundances are discussed with reference to models of gas-phase and surface chemistry.
}


   \maketitle
%

\section{Introduction}

Molecular hydrides are important in the chemistry of the interstellar medium since they often appear in
the first steps in chains of reactions that lead to other more complex species. 
The production pathways of nitrogen-bearing molecules  are  still rather uncertain since
key species, such as NH$^+$, NH, and \nh2 have not been widely observed. Even the first identified 
polyatomic interstellar molecule, ammonia (NH$_3$), has been widely observed mainly in its para symmetry
form, which leaves its formation mechanism poorly constrained in the diffuse molecular gas.
Both gas-phase chemistry and grain surface reactions have been proposed as formation mechanisms,
but 
clearly more observations are needed for a better understanding. 

Both NH and \nh2~are well known in comets (e.g. Swings et al. 1941; Meier et al. 1998; Feldman et al. 1993), 
and have been observed in stellar photospheres (e.g. Schmitt 1969; Farmer \& Norton 1989)
via their electronic, vibration-rotation, and high rotational transitions.
Interstellar NH was first detected in the interstellar medium by Meyer \& Roth (1991)
 by optical absorption spectroscopy. 
Subsequent observations by Crawford \& Williams (1997) and Weselak et~al. (2009)
have yielded six
lines of sight where column densities of both NH and H$_2$ are directly
measured. The average value of the column density ratio in these diffuse 
and translucent sightlines is $N({\rm NH})/N({\rm H}_2)=3\times 10^{-9}$.
Interstellar NH$_2$ was first observed by van Dishoeck et~al. (1993) in 
absorption towards Sgr B2 in three fine-structure components of the $para$-NH$_2$ $1_{1,0}-1_{0,1}$ 
transition with partially resolved hyperfine 
structure at frequencies 461 to 469 GHz. Further absorption lines of both
$ortho$ and $para$ forms of NH$_2$ and NH were observed through use of 
the long-wavelength spectrometer aboard the Infrared Space Observatory (ISO: Cernicharo et al. 2000; Goicoechea et al. 2004; Polehampton et al. 2007).
The ISO observations were unable to resolve the hyperfine structure
of either molecule. Ammonia, NH$_3$, was the first polyatomic molecule to be identified
in interstellar space (Cheung et al. 1968)
by means of its microwave inversion transitions.
Although NH$_3$ has been widely observed in dark clouds and star-forming regions, there have
been very few measurements of the inversion lines in the diffuse interstellar gas (Nash 1990; Liszt et al. 2006).

With the unique capabilities of \emph{Herschel}
Space Observatory  (Pilbratt et al. 2010)
transitions between 157 and 625\,$\mu$m (0.48\,--\,1.9\,THz) are available  with the Heterodyne Instrument for the Far-Infrared (HIFI; de Graauw et al. 2010)
with very high sensitivity. This allows searches for spectrally resolved, ground-state rotational transitions of  NH$^+$, NH,  \nh2, 
and \NH3~with the same instrument. Abundances of these species are key diagnostics for the nitrogen chemistry.

The most sensitive and model-independent method for measuring column 
densities of interstellar molecules is high-resolution absorption line
spectroscopy, which is being exploited by the \emph{Herschel} PRISMAS key
programme (PRobing InterStellar Molecules with Absorption line Studies).
This paper presents the first observations and analysis of nitrogen 
hydrides along the sight-line towards the massive star-forming region 
\g10~(W31C) as a part of the PRISMAS program. Observations of 
interstellar absorption towards  seven additional submillimetre-wave 
continuum sources will be presented and analysed in later publications.

The  ultra-compact HII region
\g10~in the star-forming W31 complex  is an extremely luminous submillimetre and infrared continuum source.
The source is located within the so-called 30 \kms~arm at a kinematic distance of 4.8\,kpc (Fish et al. 2003).
The gas associated directly with \g10~is detected in the velocity range 
$V_\mathrm{LSR}\approx -15$ to $+5$\,\kms, while the foreground gas along the
line of sight is detected at $V_\mathrm{LSR}\approx 5$ to 55\,\kms. 
The focus of this paper is on the diffuse interstellar gas traced by the absorption lines.

\section{Observations}
 
The observations, which took place in March 2010, are summarised
 in Table~\ref{Table: transitions}.
The 2$_0$-1$_0$  and 2$_1$-1$_1$ \NH3~transitions were observed in the same band.  
 We used the
dual beam switch mode and the wide band spectrometer (WBS) with a bandwidth of 4\,GHz and an effective spectral  resolution of 1.1\,MHz. 
The corresponding  velocity resolution 
is about 0.3\,\kms~at the higher frequencies and 0.6\,\kms~at 572\,GHz. In addition, simultaneous observations were performed using the high-
resolution spectrometer (HRS) with an effective spectral resolution of 0.25\,MHz ($\Delta \upsilon\!\sim\!0.1$\,\kms) and a bandwidth of 230\,MHz.   
Three observations were carried out with  different frequency settings of the local oscillator (LO)   corresponding to a change of approximately 15\,\kms~to
determine the sideband origin
of the lines. 
Two orthogonal polarisations were used during all observations.

 \begin{table}[\!ht] 
\centering
\caption{Observed transitions. 
}
\begin{tabular} {lrcccc} 
 \hline\hline
     \noalign{\smallskip}
Species 	& Frequency$^{a}$	&	Band$^b$ & $T_\mathrm{sys}^c$  & $t{^{\,c}_\mathrm{int}}$  &Transition   	\\    \noalign{\smallskip}
& (GHz) &  & (K) & (s)&   \\
     \noalign{\smallskip}
     \hline
\noalign{\smallskip}  
 NH$^+$ & 	1\,012.540	& 4a &  385 & 47  &   $^2\Pi_{1/2}$  $J$\,=\,3/2\,$\leftarrow$\,1/2    \\  
 NH  &	974.478	&	4a & 339 	 &30	& $N\!=\!1\leftarrow0$ $J\!=\!2\leftarrow1$  \\
$o$-NH$_2$	&	952.578	&	3b& 230  & 15& $1_{1,1}$\,--\,$0_{0,0}$ \\
$o$- NH$_3$	&	572.498		& 1b  & 83 &276  &  1$_0$-0$_0$   \\
 	&	1\,214.859	&	5a  & 1\,012   &49   & 2$_0$-1$_0$   \\
$p$- NH$_3$	 	&	1\,215.245	 &5a & 1\,012 & 49	    & 2$_1$-1$_1$   \\
    \noalign{\smallskip}
\hline 
\label{Table: transitions}
\end{tabular}
\begin{list}{}{}
\item$^{a}$The NH and \nh2~frequencies refer  to the strongest hyperfine structure component. $^{b}$HIFI consists of 7 different mixer bands and two  double sideband spectrometers. All transitions were observed in the upper sideband except NH$^+$.  $^{c}$System temperature. $^{d}$The on-source  time for each integration. 
\end{list}
\end{table} 

 \begin{figure}[\!ht] 
\centering
\resizebox{\hsize}{!}{\includegraphics{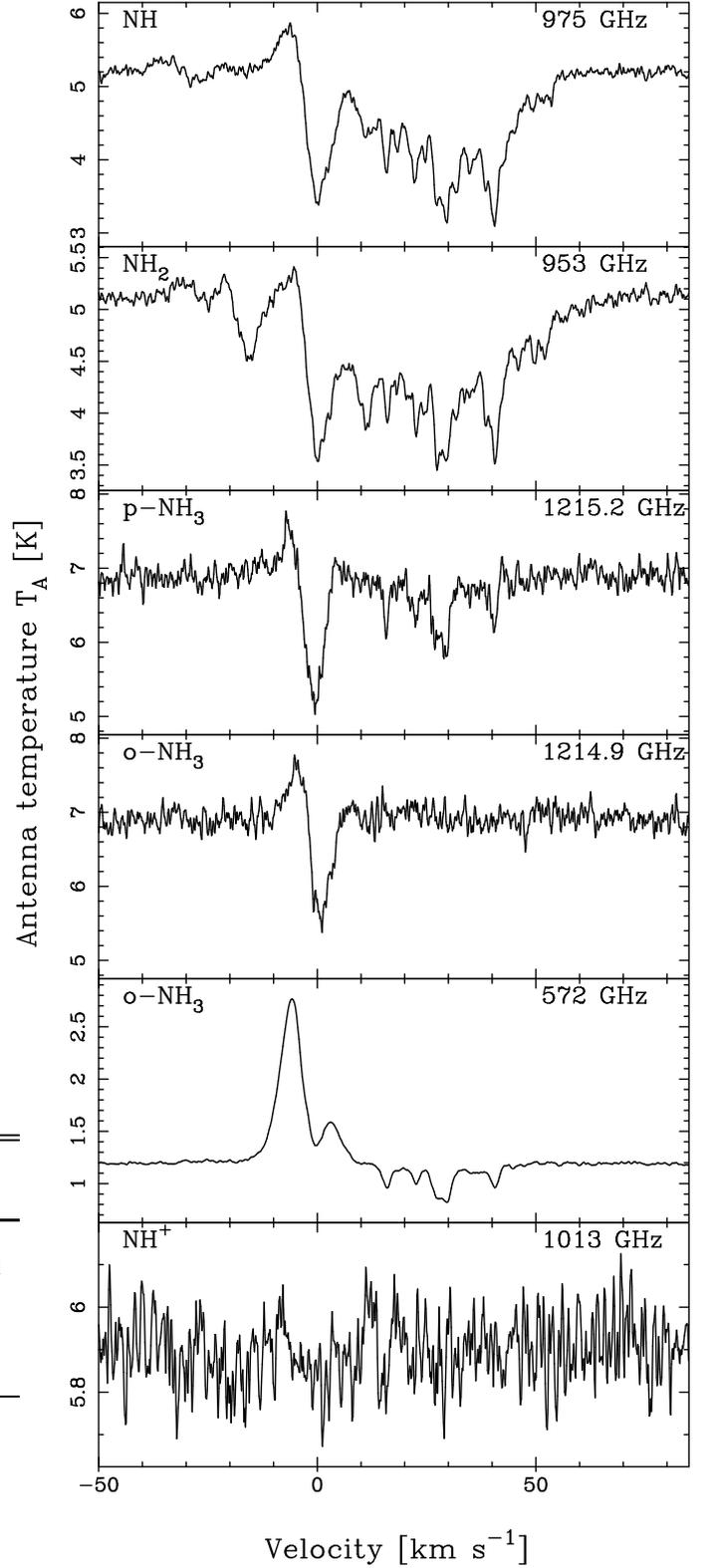}} 
 \caption{Double sideband spectra of NH, $o$-\nh2,  $o$ and $p$-\NH3, and NH$^+$  over the LSR velocity range -50 to 85\,\kms.}
 \label{Fig: All original hydrides}
\end{figure}

The half-power beam width of the telescope is  37, 22, and 17\arcsec~at 572, 953, and 1\,215\,GHz, respectively. Pointing was centered at 
R.A.\,=\,18$\fh$10$\fm$28\,$\fs$70, Dec.\,=\,$-$19$\fdg$\,55$\farcm$\,50$\arcsec$ ($J$2000). The reference beams were located 3\arcmin~on either side of the source and  
the  calibration is described by Roelfsema (2010).

We have used the standard \emph{Herschel} pipeline to Level 2 which provides fully calibrated spectra. The data were first analysed using
\emph{Herschel} interactive processing environment  
(HIPE\footnote{HIPE is a joint development by the \emph{Herschel} Science Ground
Segment Consortium, consisting of ESA, the NASA \emph{Herschel} Science Center, and the HIFI, PACS and
SPIRE consortia.}; Ott 2010)
version 2.4, and in parallel we also used the software package {\tt xs}\footnote{Developed by Per Bergman at Onsala Space Observatory, Sweden; {\tt http://www.chalmers.se/rss/oso-en/observations/}}. 
The data quality is excellent with very low intensity ripples,
two polarisations  that are in agreement to better than 10\%, and  very good agreement between the
three LO-tunings 
without contamination from the image sideband.
We thus average the three LO-tunings and both polarisations in both spectra, except for   \NH3~at 572\,GHz and NH$^+$ which display ripples in
the V- and H-polarisation, respectively, and were  therefore not included.
The resulting rms noise  is 11, 120, 50, 51, and 74\,mK for \NH3~at 572 and 1215\,GHz, for NH, \nh2, and NH$^+$, respectively.

\section{Results}

Figure~\ref{Fig: All original hydrides} shows the double sideband WBS   spectra of all observed transitions. The frequency scale is converted as customary to Doppler velocities relative to the local standard of rest ($V_\mathrm{LSR}$).
All spectra, except  NH$^+$ which is not  detected, show emission at negative velocities and a deep absorption at $-$1\,\kms~associated with \g10~as measured by observations of OH masers.
At higher velocities we see a number of absorption features   which are also seen in atomic hydrogen (Fish et al. 2003),    
HCO$^+$, HNC, HCN and CN (Godard et al. 2010). 
The only transition which does not show  absorption components from the foreground material is $ortho$-\NH3~2$_0$\,--\,1$_0$.

The double-sideband calibrated antenna temperatures  
shown in  Fig.~\ref{Fig: All original hydrides}  have to be divided by two in order to get the
correct continuum levels, $T_\mathrm{A}\mathrm{(cont)}$,  which are 0.6, 2.6. 2.6, 3.0 and 3.4\,K
for \NH3~at 572\,GHz, \nh2, NH,  NH$^+$, and \NH3~at 1\,215\,GHz,  respectively.  The  sideband gain ratio  has been shown to be close to unity in previous PRISMAS observations 
 (Neufeld et al. 2010b; Gerin et al. 2010), although departures at the 10\%\ level have been seen at some
frequencies (Neufeld et al. 2010a). We here adopt a sideband ratio of unity.

\begin{figure}[\!ht] 
\centering
\resizebox{\hsize}{!}{\includegraphics{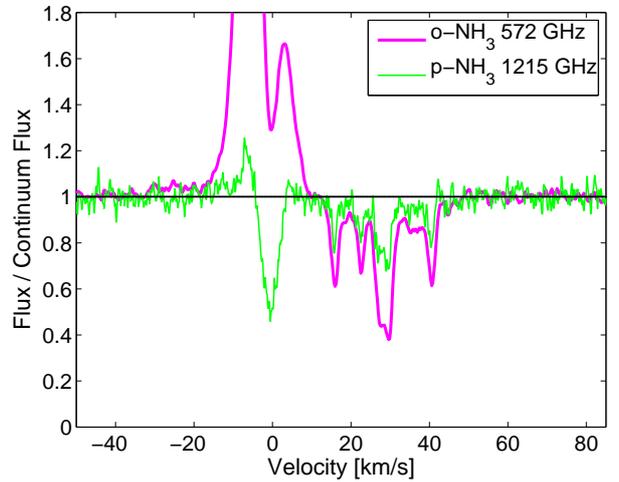}}     
\caption{Comparison of    $o$- and $p$-NH$_3$ at 572.5 and 1\,215.2\,GHz where the intensities have been normalised to single sideband continuum.}
 \label{Fig: ortho and para nh3}
\end{figure}

Figure~\ref{Fig: nh, nh2 and nh3} shows a comparison of NH, \nh2, and $para$-\NH3~where the intensities are normalised to the continuum in a single sideband as $T_\mathrm{A}$/$T_\mathrm{A}$(cont)$-$1
assuming a sideband gain ratio of unity. 
The NH and \nh2~spectra are strikingly similar, despite their complicated hyperfine structures.
The $para$-\NH3~spectra also show the same absorption pattern, although not as strongly.  
Figure~\ref{Fig: ortho and para nh3} shows a similar comparison of $ortho$- and $para$-\NH3. Here, too, we see the same absorption pattern in both species,  
and strong emission and self-absorption in the $ortho$-line from the source itself.
The strongest velocity components in the foreground material lie at $V_{\mathrm{LSR}}=$ 
16, 22.5, 27.5, 30 and 40.5\,\kms.

The line opacities are calculated as \mbox{$\tau\!=\!-\ln{(T_\mathrm{line}/T_\mathrm{A}\mathrm{(cont)}-1)}$}.
To be conservative and not include absorption from the source itself, we use the velocity interval \mbox{11\,--\,54\,\kms}.
The total integrated opacities are $\int \tau dV=$ 20, 16, 9, and 3.4\,\kms~for NH, \nh2, \NH3~at 572 and 1\,215\,GHz, respectively.

\begin{figure}[\!ht] 
\centering
\resizebox{\hsize}{!}{\includegraphics{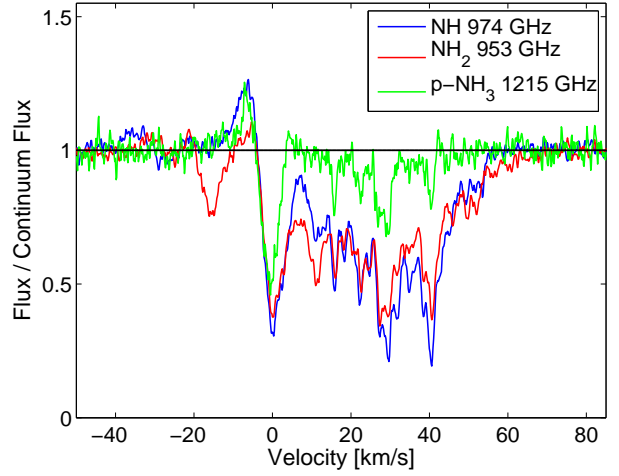}} 
\caption{Comparison of  NH, $o$-NH$_2$ and $p$-NH$_3$ at 1\,215.2\,GHz where the intensities have been normalised to single sideband continuum.}
 \label{Fig: nh, nh2 and nh3}
\end{figure}

Even though unsaturated absorption lines provide straightforward determinations of opacity and column densities, the numerous hyperfine-structure components and velocity features complicate the 
analysis of both NH and  \nh2. 
The  spectroscopic properties of NH and \nh2~have been studied extensively  and 
references, rest frequencies and molecular properties for NH and  \nh2~are listed in the Cologne Database for Molecular Spectroscopy\footnote{\tt{http://www.cdms.de}} 
(CDMS, M\"uller et~al. 2005), and  \NH3~in Jet Propulsion Laboratory\footnote{\tt{http://spec.jpl.nasa.gov}} (JPL, Pickett et~al. 1998).
 
In order to estimate the total column density from the integrated opacity, it is necessary to 
correct for the population of molecules in unobserved excited levels and to quantify any possible
effect of stimulated emission. We have done this through non-equilibrium excitation calculations
of integrated opacity
with an enhanced version of the  {{\tt RADEX}} code\footnote{The published 
version will soon include these enhancements, see
{\tt http://www.sron.rug.nl/{$\sim$}vdtak/radex/index.shtml}}  (van der Tak et~al. 2007).
Where collision rates are unknown, we have made guesses scaled in proportion to radiative line 
strengths. 
We assume  diffuse molecular cloud conditions with $T_\mathrm{k}$\,=\,30\,K and $n$(H$_2$)\,=\,500\,\cmcub.  
The background radiation is a model of the average Galactic background radiation in the solar neighbourhood plus the 
cosmic microwave background radiation. 
The excitation of the hydride molecules in diffuse gas 
is dominated by the background continuum radiation and not by collisions. The resulting excitation temperatures of the observed submillimetre 
transitions are typically \mbox{4\,--\,5\,K}, which
are small enough compared to $h\nu/k$ that no correction for emission is required. 
All the optical depths of the hyperfine components that contribute to the line are summed. We have
treated $ortho$- and $para$-forms of molecules together in the expectation that their 
relative abundances are fixed by the chemical formation process. Plausible rates of formation
and destruction are included explicitly in the excitation calculation. Thus we derive total column densities as summarised in  
Table~\ref{Table: columns}. The analysis is complicated for NH$_3$
by the existence of metastable excited states, especially in the $para$ species. The entry for NH$_3$ 
in Table~\ref{Table: columns} is the sum of the measured column density in $J_K\!=0_0$, the measured
column density in the lower inversion sublevel of $1_1$, and calculated values in the upper inversion
sublevel of $1_1$ and in both sublevels of $2_2$. All other states are expected to have negligible
populations. The uncertainties of measurement and in the excitation analysis do not fully constrain
the $ortho$/$para$ ratio in NH$_3$, but it appears that the effective formation temperature must
be greater than 20\,K,   otherwise  the strength of the 1\,215\,GHz line cannot be explained.

\begin{table}[\!ht] 
\centering
\caption{Total column densities, $N$,    and abundances with respect to hydrogen, $X$.
}
\begin{tabular} {lcc} 
 \hline\hline
     \noalign{\smallskip}
Species $x$		  	&$N^a$ & $X$\,=\,$N_x/N_H$	\\    \noalign{\smallskip}
  	& (\cmsq) & \\
     \noalign{\smallskip}
     \hline
     \noalign{\smallskip}
NH  &1.5$\times$10$^{14}$	& 5.6$\times$10$^{-9}$  \\  
NH$_2$	 	  	&8.0$\times$10$^{13}$&3.0$\times$10$^{-9}$  \\ \noalign{\smallskip}   
NH$_3$	 	 	  &	8.7$\times$10$^{13}$ & 3.2 $\times$10$^{-9}$	\\  
NH$^+$  	  	& $\lesssim2\times$10$^{13}$  &  $\lesssim8\times$10$^{-10}$	  \\
   \noalign{\smallskip}
\hline 
\label{Table: columns}
\end{tabular}
\begin{list}{}{}
\item$^{a}$Calculated in  the velocity interval \mbox{11\,--\,54\,\kms}.
\end{list}
\end{table} 

 We note that the three neutral nitrogen hydrides have comparable abundances: NH is approximately
twice as abundant as \nh2~and \NH3. To estimate the  abundance with respect to hydrogen 
we use $N_H$\,=\,2.7$\times$10$^{22}$\,\cmsq~as total column density of   hydrogen in all forms    (Neufeld et~al. 2010b).
This value is consistent with the estimated atomic hydrogen column density in this line-of-sight,  $N$(H)\,=\,1.2$\times$10$^{22}$\,
\cmsq~(Godard et~al. 2010)  
inferred from absorption profiles observed with the VLA   (Fish et~al. 2003), 
and implies that the absorbing 
material is mostly molecular even if there most likely are large variations between the individual velocity components. 
Our estimate of the abundance of NH relative to total hydrogen is somewhat higher
than the NH/H$_2$ ratio of $3\times 10^{-9}$ for six lines of sight
where both are observed directly.
The resulting abundances are found in  Table~\ref{Table: columns}.
The non-detection of NH$^+$ is used to put an upper limit on its column density and abundance 
(3$\sigma$).
A previous limit on NH$^+$ towards $\zeta$~Oph from ultraviolet observations, 
$N({\rm NH}^+)/N_{\rm H}<1\times 10^{-6}$, is 
orders of magnitude less sensitive than our limit (Snow 1979; de Almeida \& Singh 1982).

\section{Discussion}

The production of NH and \nh2~by purely gas-phase processes is problematic since their formation is inhibited by a lack of a sufficient source of N$^+$. The atomic ion can be formed by 
cosmic ray ionization of N or by reaction of He$^+$ with N$_2$ or CN, which were formed by
neutral-neutral reactions.The standard gas-phase ion-molecule chemistry of nitrogen hydrides is then initiated by the reaction N$^+$+H$_2\!\rightarrow$\,\,NH$^+$ + H, because H$_3^+$ does not
react rapidly with N.
Typical steady state  \emph{dark cloud} models ($n$\,=\,1$\times$10$^{3}$\,--\,5$\times$10$^{4}$\,\cmcub, 
$T$\,=\,10\,--\,40\,K, $A_\mathrm{V}\gtrsim10$),   predict an \nh2~abundance of \mbox{(1-10)$\times$10$^{-8}$}, an	 NH abundance 10 times lower, 
and an \nh2/\NH3~ratio of  \mbox{0.3\,--\,1.5} for a wide range of assumptions  
(e.g. Millar et~al. 1991; Langer \& Graedel 1989), but these are not directly applicable
to diffuse molecular gas. 

Previous observations of NH in diffuse clouds were taken to imply that 
NH is produced on grain 
surfaces (Mann \& Williams 1984; Wagenblast et~al. 1993; O'Neill et~al. 2002).
Such models, however, often predict up to 1\,000 times more \NH3~than \nh2~(Hasegawa \& Herbst 1993).
An additional source of NH and \nh2~is fragmentation of ammonia by photodissociation in diffuse clouds. Models of ultraviolet illuminated PDRs      predict 
\nh2/\NH3$<$1 and \NH3/NH$<$1
in regions with extinction \mbox{$A_v\lesssim5\,\mathrm{mag}$} (e.g. Sternberg \& Dalgarno 1995).
The abundance patterns that we see in diffuse molecular gas are clearly different from those in
Sgr B2, where \NH3/\nh2/NH $\sim 100/10/1$ and the fractional abundance of NH is 
a few times $\times$10$^{-9}$ (Cernicharo et~al. 2000; Goicoechea et~al. 2004).

We have used two different approaches to chemical models in order to compare the nitrogen-hydride
abundances with theory.
First, we constructed a pseudo-time-dependent model  with constant physical conditions      
including both gas-phase and surface chemistry 
(Hasegawa et~al. 1992)
with $A_V$\,=\,2 and 3, $n_\mathrm{H}$\,=\,500 and 5\,000\,\cmcub, $T_\mathrm{gas}$\,=\,30 and 50\,K, and $T_\mathrm{dust}$\,=\,10\,K.  
The ultraviolet  field is
1\,$G_0$, and the cosmic ray ionization rate is $\zeta$\,=\,1.3$\times$10$^{-17}$\,s$^{-1}$.    
The resulting abundances are found in Fig.~\ref{Fig: Hassel Model} (online material).
Gas-phase chemistry alone does not match the observed abundances, failing by factors of 10 to 100
for NH and NH$_2$. Inclusion of processes on dust surfaces increases the abundances of the three
neutral hydrides to 10$^{-8}$\,--\,10$^{-9}$, but fails to match the high NH/NH$_3$ ratio.
If the dominant source of \NH3~is association on surfaces  and if the $ortho$/$para$ ratio is fixed at a surface temperature below  20\,K, then the 1\,215\,GHz $para$ line should be much weaker relative to the $ortho$  572\,GHz line than is observed. The observed line ratio is perhaps somewhat better explained by a high formation temperature, which points to exoergic gas-phase formation of \NH3.

A second approach is to construct models of weak photon-dominated regions (PDR) through use of the
Meudon PDR code (Le Petit el~al. 2006; Goicoechea \& Le Bourlot 2007), 
which is a steady-state, depth-dependent model with pure gas-phase chemistry.
We use parameters appropriate
for diffuse clouds, $UV$\,=\,0.5\,--\,5\,$G_0$,  $n_\mathrm{H}$\,=\,100\,--\,1\,000\,\cmcub, $\zeta$\,=\,10$^{-17}$\,--\,10$^{-15}$\,s$^{-1}$,  $A_\mathrm{V}\!<5$ and  illuminate one side only. The   results of the best fitting model are found in Fig.~\ref{Fig: PDR Model} (online material).
Although the resulting relative abundances of nitrogen-hydrides approach the observed ratios inside the cloud, the abundances relative to hydrogen are still too low.

\section{Conclusions}

Our first detections of spectrally resolved rotational transitions of nitrogen hydrides in the interstellar medium show remarkable similarities of line profiles and abundances towards one 
background source. The formation mechanisms are, however, difficult to explain.
Neither standard gas-phase nor grain-surface chemistry can fully explain our observations, which
may suggest that important reactions have been overlooked in the chemistry of interstellar
nitrogen. 
Further analysis of the line
profiles will yield abundance ratios in separate velocity components and will allow the 
nitrogen species to be compared directly with tracers of molecular hydrogen like CH and HF. 
Our sample will also be enlarged with seven additional sight-lines in the PRISMAS project. The
chemical models will be refined and the possible role of turbulent dissipation regions will
be assessed.

\begin{acknowledgements}
HIFI has been designed and built by a consortium of institutes and university departments from across
Europe, Canada and the United States under the leadership of SRON Netherlands Institute for Space
Research, Groningen, The Netherlands and with major contributions from Germany, France and the US.
Consortium members are: Canada: CSA, U.Waterloo; France: CESR, LAB, LERMA, IRAM; Germany:
KOSMA, MPIfR, MPS; Ireland, NUI Maynooth; Italy: ASI, IFSI-INAF, Osservatorio Astrofisico di Arcetri-
INAF; Netherlands: SRON, TUD; Poland: CAMK, CBK; Spain: Observatorio Astronómico Nacional (IGN),
Centro de Astrobiología (CSIC-INTA). Sweden: Chalmers University of Technology - MC2, RSS \& GARD;
Onsala Space Observatory; Swedish National Space Board, Stockholm University - Stockholm Observatory;
Switzerland: ETH Zurich, FHNW; USA: Caltech, JPL, NHSC.
CP and JHB acknowledge generous support from the Swedish National Space Board. 
JC and JRG thanks spanish MICINN for funding support under projects AYA2009-07304 and CSD2009-00038
M.S. acknowledges support from grant N\ 203\ 393334 from Polish MNiSW.
\end{acknowledgements}

  {}
 
\Online
\appendix
 
\section{Figures}
 
\begin{figure*} 
\centering
\includegraphics[scale=0.65, angle = 90]{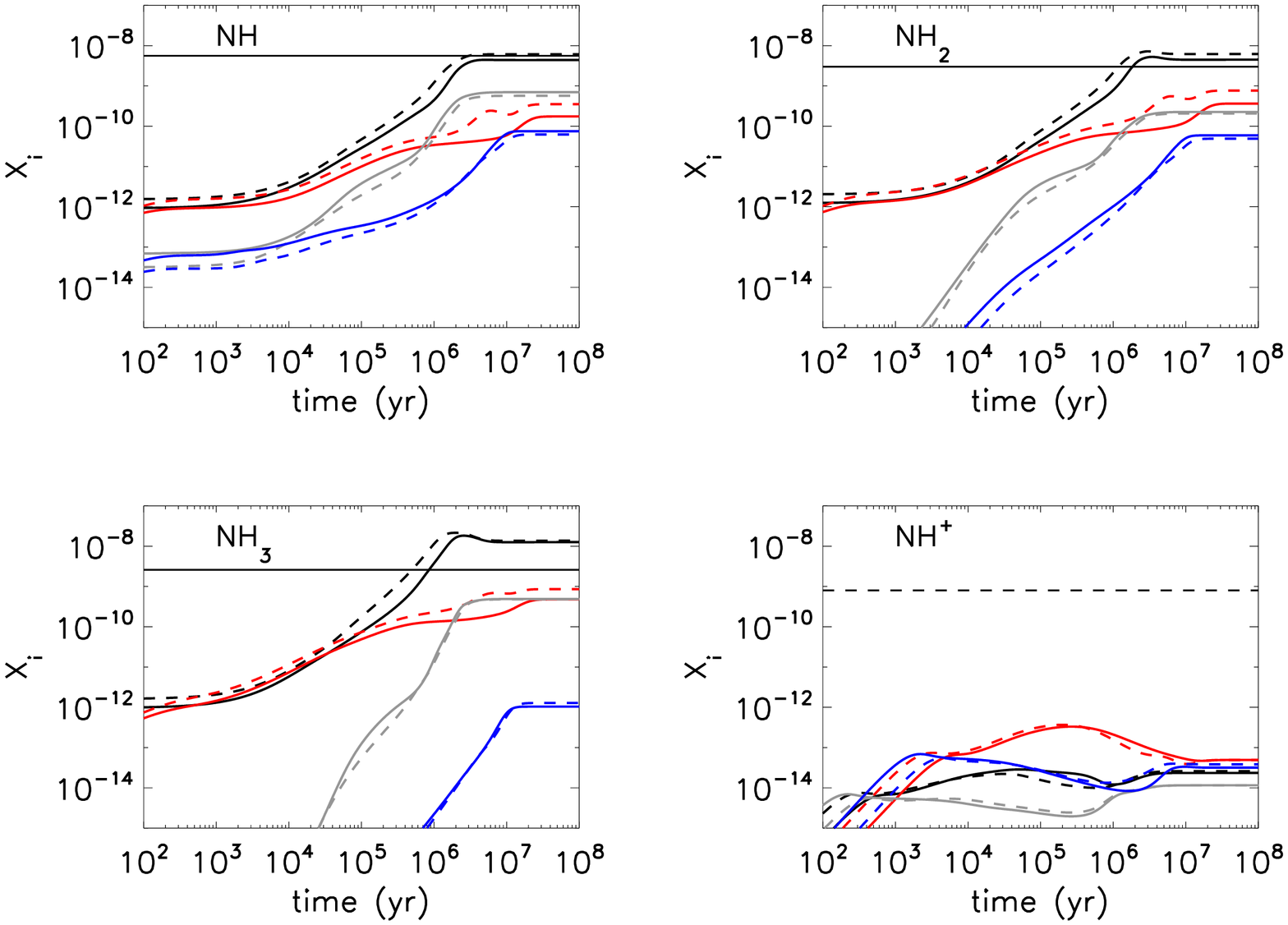}
\caption{A pseudo-time-dependent model  with constant physical conditions      taking into account both gas-phase and  grain surface chemistry.
For all lines: $T_\mathrm{dust}$\,=\,10\,K and the solid/dashed lines show $T_\mathrm{gas}$\,=\,30 and 50\,K, respectively. 
The black thin lines show the observed abundances, and the dashed black line the NH$^+$ upper limit.
\emph{Black} line: gas phase \&  grain surface chemistry using $A_V$\,=\,3, $n_\mathrm{H}$\,=\,5000\,\cmcub. 
\emph{Red} line: gas phase \& grain surface chemistry using $A_V$\,=\,2, $n_\mathrm{H}$\,=\,500\,\cmcub.
\emph{Grey} line: only gas phase chemistry using $A_V$\,=\,3, $n_\mathrm{H}$\,=\,5000\,\cmcub.
\emph{Blue} line: only gas phase chemistry using  $A_V$\,=\,2, $n_\mathrm{H}$\,=\,500\,\cmcub.
}
 \label{Fig: Hassel Model}
\end{figure*}

 \begin{figure*}[\!ht] 
\centering
\includegraphics[scale=0.9, angle =-90]{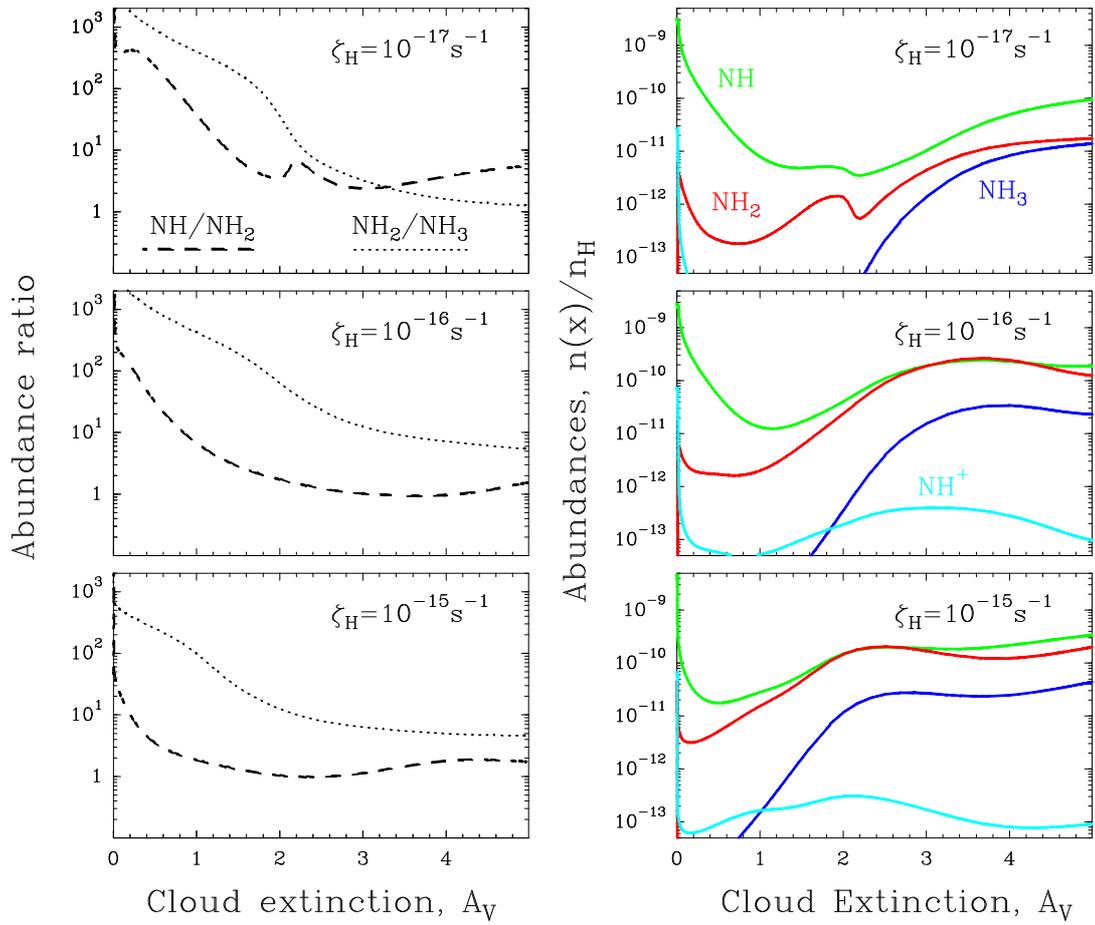}
\caption{The cloud depth-dependent Meudon PDR model for pure gas-phase chemistry using $n_\mathrm{H}$\,=\,1\,000\,\cmcub,  $\zeta$\,=\,10$^{-17}$\,--\,10$^{-15}$\,s$^{-1}$, and a UV-field of  10 times the interstellar radiation
field in Draine units ($\sim$5\,$G_0$). The left panel shows the relative abundances of NH/\nh2~and \nh2/\NH3, and the right panel show the absolute abundances.}
 \label{Fig: PDR Model}
\end{figure*}

 \end{document}